\documentclass[11pt,twoside]{article}
\usepackage{./asp2014}

\usepackage{graphics,graphicx}
\usepackage{ulem}

\aspSuppressVolSlug
\resetcounters

\begin{document}
\title{Resolving the Radio Complexity of EXor and FUor-type Systems with the ngVLA}
\author{Jacob Aaron White$^{1}$, Marc Audard$^{2}$, P\'eter \'Abrah\'am$^{1}$, Lucas Cieza$^{3}$, Fernando Cruz-S\'aenz de Miera$^{1}$, Michael M. Dunham$^{4}$, Joel D. Green$^{5}$, Manuel G\"udel$^{6}$, Nicolas Grosso$^{7}$, Antonio Hales$^{8}$, Lee Hartmann$^{9}$, Kundan Kadam$^{1}$, Joel H. Kastner$^{10}$, \'Agnes K\'osp\'al$^{1}$, Sebastian Perez$^{11}$, Andreas Postel$^{2}$, Dary Ruiz-Rodriguez$^{10}$, Christian Rab$^{12}$, Eduard I. Vorobyov$^{13,14}$, \& Zhaohuan Zhu$^{15}$}
\affil{$^{1}$Konkoly Observatory, Hungarian Academy of Sciences, Budapest, Hungary}
\affil{$^{2}$Department of Astronomy, University of Geneva, Versoix, Switzerland}
\affil{$^{3}$Facultad de Ingenier\'ia y Ciencias, Univ. Diego Portales, Santiago, Chile}
\affil{$^{4}$Dept. of Physics, State University of New York at Fredonia, Fredonia, USA}
\affil{$^{5}$Space Telescope Science Institute, Baltimore, USA}
\affil{$^{6}$Department of Astrophysics, University of Vienna, Vienna, Austria}
\affil{$^{7}$Aix Marseille Univ, CNRS, CNES, LAM, Marseille, France}
\affil{$^{8}$Joint ALMA Observatory, Santiago, Chile}
\affil{$^{9}$Dept. of Astronomy, University of Michigan, Ann Arbor, USA}
\affil{$^{10}$Center for Imaging Science, Rochester Inst. of Technology, Rochester, USA}
\affil{$^{11}$Departamento de Astronom\'ia, Universidad de Chile, Santiago, Chile}
\affil{$^{12}$Kapteyn Astronomical Inst., Univ. of Groningen, Groningen, NL}
\affil{$^{13}$University of Vienna, Department of Astrophysics, Vienna, Austria}
\affil{$^{14}$Research Inst. of Physics, Southern Federal Univ., Roston-on-Don, Russia}
\affil{$^{15}$ Dept. of Physics and Astronomy, University of Nevada, Las Vegas, USA}

\paperauthor{Jacob Aaron White}{jacob.white@csfk.mta.hu}{0000-0001-8445-0444}{Konkoly Observatory}{}{Budapest}{}{1121}{Hungray}

\begin{abstract}

Episodic accretion may be a common occurrence in the evolution of young pre-main sequence stars and has important implications for our understanding of star and planet formation. Many fundamental aspects of what drives the accretion physics, however, are still unknown. The ngVLA will be a key tool in understanding the nature of these events. The high spatial resolution, broad spectral coverage, and unprecedented sensitivity will allow for detailed analysis of outburst systems. The proposed frequency range of the ngVLA allows for observations of the gas, dust, and non-thermal emission from the star and disk.
 
\end{abstract}

\section{Young Outburst Systems}

Understanding the underlying mechanisms that cause the observed outbursts of young stellar objects (YSOs) is important not only for building a complete picture of stellar formation, but also for the potential implications on the planet formation process. These outbursts offer a potential solution to the \emph{luminosity problem}, i.e., the discrepancy between the mass accretion rates inferred from protostellar luminosities (10$^{-8}$ M$_\odot$ yr$^{-1}$) and the accretion rates predicted by cloud core collapse models (10$^{-6}$ M$_\odot$ yr$^{-1}$). The observed outbursts may be due to \emph{episodic accretion} of disk material onto the star, which consists of a sudden increase in the accretion rate from 10$^{-8}$ M$_\odot$ yr$^{-1}$ in quiescence up to as high as 10$^{-4}$ M$_\odot$ yr$^{-1}$ during an outburst. The relatively short $\sim1-100$ yr cumulative duration of the outbursts, compared to the quiescent phase, provides a potential solution to the luminosity problem \citep[e.g.,][]{dunham12}.

The study of these outburst phenomena began with FU Orionis \citep{herbig66}, which in 1937 showed an increase of 5 magnitudes in the V band and has since been slowly fading. Outbursts are now broadly categorized into two groups: FUors and EXors, named after their prototypes FU Orionis and EX Lupi, respectively \citep[see][for a recent review of these types of objects]{audard14}. FUors undergo large outbursts which can last 100s of years, while EXors exhibit smaller and shorter lived outbursts ($\sim1$ yr). Indeed, there are objects that do not fit clearly into either classification (FUor-like objects) as they can show similar spectroscopic characteristics as a FUor without showing other evidence of an outburst. This leads to the possibility that EXors and FUor represent different stages of evolution of the same object. \citet{cieza18} showed that EXors, for example, have lower disk masses than FUors, indicating they may represent a later stage of disk evolution. Although over the past 50 years EX/FUors have been observed with broad spectral coverage and modeled with increasing high levels of detail, there are still many open questions on their formation, frequency of occurrence, and impact on circumstellar environments.

\emph{What are the physical processes that cause these types of events?} Despite being thoroughly studied, the triggering mechanisms of these outbursts are still not known. It is also not clear if the same mechanisms are responsible for both EXor and FUor-type outbursts. The ngVLA will be instrumental in addressing this issue by: 

\begin{enumerate}

\item Accurately measuring disk masses. Knowing disk masses is crucial for understanding the FUor phenomenon, as gravitational instability may play an important role in triggering of the outbursts \citep[e.g.,][]{vorobyov05,Bae2014}. The dust disks are optically thick and have a small angular size, which complicates mass estimates \citep[e.g.][]{liu17, hales18}. However, the unprecedented angular resolution of the ngVLA, in combination with its spectral coverage at longer wavelengths where the dust can be optically thin, will enable more precise dust mass estimates. In addition, higher resolution observations of cold gas will provide further constraints on the total disk mass.

\item Dynamically constraining stellar masses. 
As the luminosity of EX/FUor systems can be dominated by the disk, it can be difficult to accurately determine the mass of the protostar \citep[e.g., V883 Ori has an expected photospheric luminosity of $\sim6$ L$_\odot$ and a bolometric luminosity of $\sim400$ L$_\odot$;][]{cieza16}. Moreover, the inferred disk inclination affects the inferred disk luminosity and accretion rate. High spatial resolution spectroscopy with the ngVLA will allow for dynamical constraints on both the stellar mass and inclination through the analysis of the Keplerian rotation of the gas disk \citep[e.g.,][]{czekala15}.

\item Determining the strength of turbulence in the inner disk. High disk turbulence has been suggested as the reason for the loss of angular momentum needed to trigger an enhanced accretion event. \citet{flaherty17} analyzed the DCO$^+$ in HD 163296's outer disk and found little turbulence, going against the theoretical predictions. This could be due to HD 163296's more evolved state or a diminished role of turbulence-driven accretion. The ngVLA will help in understanding the role of turbulence in the innermost parts of disks, where the burst is expected to be triggered, and how it can cause eruptive events.

\item Observing planet-disk-star interactions. The mass exchange between a young protoplanet and the star in the innermost parts of the disk may trigger accretion outbursts \citep{nayakshin12}. \citet{ricci18} estimated that the ngVLA will be able to characterize gaps and asymmetries caused by a super-Earth planet at close radii ($<$5 au). At those distances, it will be possible to see changes in the disk material distribution in a few years, thus making important mass transfer rate predictions in the innermost regions of the disk.

\item Assessing the impact of stellar encounters and multiplicity. Close stellar encounters either with an external star or a companion may trigger accretion outbursts \citep[e.g.,][]{bonnell92, pfalzner08}. The spatial arrangement of the outburst star and the inferred intruder is sometimes not consistent with the timescales of the outburst \citep{liu17}, suggesting another companion in the inner unresolved disk may be present. The ngVLA will search for close-orbit companions and assess the role of companions/intruders as outburst triggers.
\end{enumerate}

\emph{What are the implications of these outburst events on the origin of our Solar System and for other planetary systems?} The location of snow-lines (or ice-lines) in a planet forming disk is important as an evolving snow-line can greatly impact the abundance of solid material throughout the disk and affect grain growth. ALMA observations of V883 Ori show the water snow-line at a radial distance of 42 au, in contrast with the expected distance of $<$5 au for a solar-like star \citep{cieza16}. Therefore, if these eruptive events are a common occurrence in the evolution of protoplanetary disks then their affects on the relocation of snow-lines, and how this can spur grain growth, must be considered in planet formation models.

In this chapter we explore how the proposed next generation Very Large Array (ngVLA) will be an invaluable tool for studying FUor and EXor type protostars. In Sec.\,2, we highlight key areas that still remain largely unexplored or unconstrained with current capabilities. In Sec.\,3, we show the uniqueness of the ngVLA's capabilities to study EX/FUors. In Sec.\,4, we highlight synergies at other wavelengths.

\section{Exploring Accretion Mechanisms with the ngVLA}

Current millimeter and radio observatories do provide valuable insight into EX/FUor systems \citep[e.g.,][]{kospal17, liu17, cieza18, hales18}, but the major limitation in these studies is typically the resolving capabilities of the observatory (e.g., $\sim20$ mas or $\sim8$ au at the Orion Nebula). When the circumstellar disk around a given EX/FUor system is unresolved, or only marginally resolved, it can be difficult (if not impossible) to test many of the competing models for accretion-driven outbursts. The ngVLA will provide a major improvement in the resolving capabilities of current facilities that operate in the $1-116$ GHz frequency range, where the inner disk may be transitioning to an optically thin regime.

There will indeed be many properties of EX/FUors, and circumstellar environments in general, that the ngVLA will be able to study. Some of the key areas are highlighted below.

\subsection{Disk Morphology}

The proposed 1000 km maximum baselines of the ngVLA will yield a resolution of 0.7 mas at 93 GHz \citep{selina18}. This resolution corresponds to spatial scales of 0.25 and 0.10 au for the two prototypical outburst systems, FU Orionis and EX Lupi, respectively. For EX Lupi, this spatial scale is slightly smaller than the $\sim0.12$ au corotation region \citep{sicilia15} and smaller than the $\sim0.4$ au inner dust-free zone \citep{abraham09}. The proposed largest recoverable scales of >20 arcsec \citep{selina18} will correspond to spatial scales of 1000s of au for FU Orionis and EX Lupi, allowing for all material with sufficient surface brightness to be detectable in a single pointing. The high level of spatial resolution will enable a detailed analysis of the morphological structure of the material in the disks.

Many of the proposed drivers of accretion in  EX/FUor-type objects will have distinct morphological features that can be distinguished with the ngVLA. These features are caused by mechanisms such as gravitational instabilities, disk fragmentation, circumbinary disk interactions, and funnel flows \citep[e.g.,][]{armitage01, zhu09, hartman16}. These features may be present both in the inner optically thick and the outer optically thin parts of the disks. However, current resolving capabilities make it difficult to confirm or reject the presence of any substructure in EX Lupi and FU Orionis \citep{liu17, hales18}.
 
The mm-cm spectral index of an EX/FUor-type disk will yield insight into the opacity and size distribution of the grains. An optically thick inner disk may be common, but the dust opacity may transition to an optically thin regime at the lower frequency range of the ngVLA. Indeed, it is likely that non-thermal emission will be significant at lower frequencies, but the high resolution capabilities should enable the thermal and non-thermal emission components to be spatially separated. Constraints on the size distribution (q) of solids can be made when a disk is optically thin. By measuring q, the grain growth, total dust mass, and mass distribution can be much more accurately constrained than is possible with current observations. 

The resolution and sensitivity of the ngVLA will allow for the first tests of time-dependent evolution of disk structure. There are many mechanisms that could cause significant changes to the morphological structure of the disk on timescales that can be probed with the ngVLA. For example, if a close-in companion were present, it could open a gap in the disk or cause significant spiral structure to form \citep[e.g.][]{paardekooper04, dong15, ricci18}. This disturbance to the disk could cause an infall of material which could further drive accretion. The movement of this material could potentially be traced over a few years. Changes over much shorter timescales would also be observable. Many disk models (e.g., the model discussed in Sec.\,3) have disk dead zones at separations of $\sim0.5$ au from the star. The orbital period of material at this distance would be on the order of months, and multiple observations over an ngVLA observing semester could probe morphological changes in the disk over the timescale of a single orbit. Low levels of variability in EXor/FUor-type stars have been observed at optical wavelengths (i.e., significantly lower than the outbursts themselves). If this variability is correlated with changes in the disk structure, then it could be potentially observed over the course of a single observation with the ngVLA.

\subsection{Disk Chemistry}

Winds and outflows are fundamental ingredients of star formation and protoplanetary disk evolution \citep{williams11}. Thus, it is not surprising that YSOs undergoing episodic accretion, such as FUors, also show high levels of outflow activity. These outflows are responsible for carrying away some of the angular momentum, allowing circumstellar matter to accrete onto the forming star. Moreover, previous studies suggest that outflows could be an effective way to disrupt the disk and envelope at early evolutionary stages \citep{ruizrodriguez17a, ruizrodriguez17b}. Accordingly, it is imperative to investigate the relation between accretion and outflows in young stellar systems, so as to disentangle the physical mechanisms that dictate the low mass star formation efficiency in turbulent clouds \citep{quillen05, krumholz12} and efficient transport of angular momentum that permits accretion onto the star.

In addition, strong X-ray emission can be produced during FUor eruptions and rapid YSO accretion episodes more generally \citep{audard14}, either by accretion shocks \citep{grosso10,teets12}, magnetic reconnection \citep{kastner04}, or both. It has been known that X-ray radiation is a critical ionizing source in the surface disk layers at distances of $\sim10-100$ au \citep{glassgold12} and plays an important role in the dispersal of disks at pre-main sequence ages $>$1 Myr \citep{gorti15}. This is due to the large penetration depth of X-rays in the disk-envelope system, where they can ionize H$_{2}$ molecules as well as abundant atoms, such as He, C, N, and O. Hence, X-ray irradiation plays a role in the formation and position of molecular disk snow lines and dead zones, thereby ultimately determining the semimajor orbits, masses, and compositions of any resulting planets \citep{oberg16,cridland17}. The resulting enhanced molecular ionization leads to rich chemistry that can be probed by tracers such as HCN and HCO$^{+}$ \citep[][and refs.\ therein]{kastner08}.

Spatially resolved observations of outflow/disk systems with the ngVLA in the wide range of chemical tracers accessible to its Band 6 (70-116 GHz) will allow us to distinguish between chemistry induced by outflow energetics (shocks) and radiation (X-rays). Specifically, sub-arcsecond resolution mapping of lines of SO, SiO, CH$_{3}$OH, and H$_{2}$CO will provide diagnostics of shock-generated chemistry within FUor outflows. Observations of the same systems in HCO$^+$ will then serve to probe the presence and extent of molecular ionization, via spatial analysis of the HCO$^+$/$^{13}$CO surface brightness ratio. In addition, using $^{12}$CO,  $^{13}$CO, C$^{18}$O line data, and adopting standard methods for correcting optical depth effects \citep[e.g.,][]{dunham14, ruizrodriguez17a, ruizrodriguez17b}, provides estimates of the mass and kinematic properties of the revealed outflows. In parallel, maps of HCN and CS emission will trace the dense gas within the disk and disk-outflow interface regions. 

\subsection{Magnetic Fields \& Polarization}

Some of the most promising outburst theories for EXors rely on the role of the stellar magnetic field. \citet{armitage16} predicted that the outbursts could be explained by changes in the polarity and strength of the stellar magnetic fields at the kG-level. In a competing scenario, \citet{dangelo12} proposed an instability which can lead to quasi-periodic oscillations in the inner disk and associated recurrent outbursts. This instability can occur when the accretion disk is truncated close to the corotation radius by the strong magnetic field of the star. A confirmation of the magnetic field strength would give strong support to these models, and may constrain the cause of outbursts to magnetic phenomena in young stars, a novel result. In principle, if kG-level stellar magnetic fields were present in a given system then there could be significant emission above the photospheric flux. For example, strong magnetic fields could induce synchrotron or free-free emission within a few stellar radii. The flux from this could be easily spatially separated from the inner edge of the disk and could provide constraints on the strength of the stellar magnetic fields.

The effects of magnetic fields within the disk itelf can be traced through polarization. For example, small dust grains can align with magnetic fields and cause polarization \citep[e.g.,][]{hull14}, allowing for another probe of the grain size distribution throughout the disk. Bright molecular lines can also be linearly polarized by magnetic fields within the disk \citep[e.g., the Goldreich-Kylafis effect;][]{forbich08}.

The large outflow jets discussed in Sec.\,2.2 can generate polarized synchrotron emission (see e.g., Galv\'an-Madrid et al. science chapter, for a detailed overview of jets in disks). These jets have been observed in both low and high mass protostellar systems and may provide another tool in studying how accretion events can vary as a function of stellar mass. The synchrotron emission associated with these jets can lead to polarization at the lower frequency range of the ngVLA. As these jets can be much more extended from the circumstellar disk, it should be straightforward to spatially separate their emission from any disk and other polarized or non-thermal emission.

\section{Uniqueness to ngVLA Capabilities}

\begin{figure}
\centering
\includegraphics[width=\textwidth]{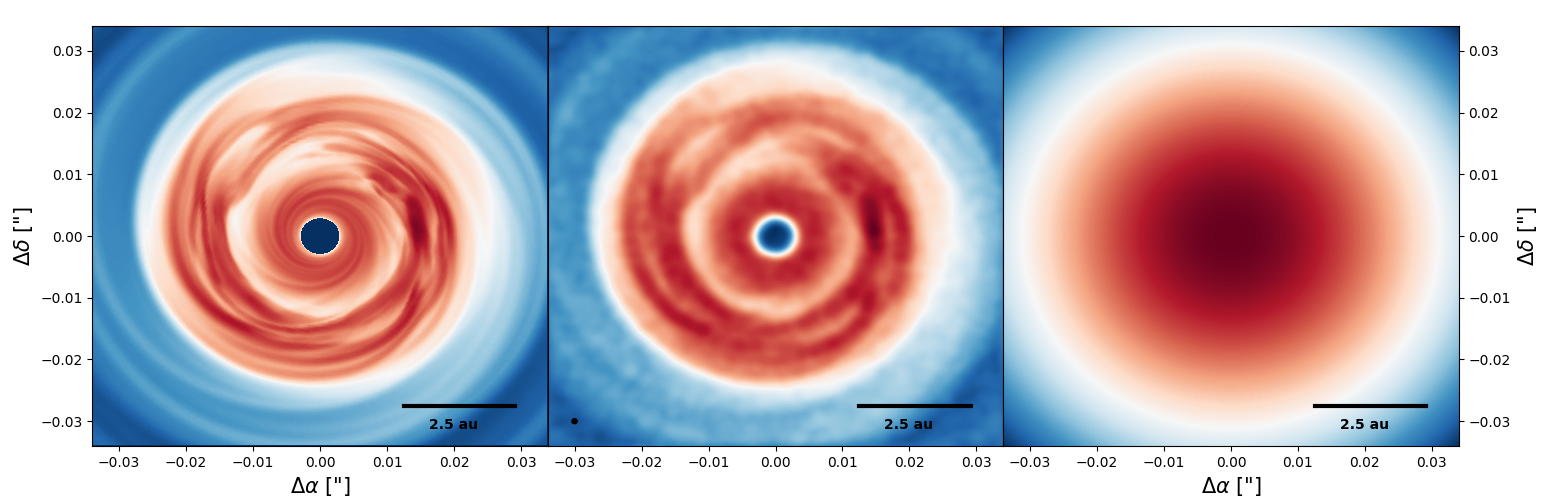}
\caption{Simulated observations with a $10\times10$ au disk model (described in Section 3), 0.7 mas 100 GHz ngVLA observations, and 40 mas 100 GHz ALMA observations. A 2.5 au scale is given in the bottom right of each panel. The synthesized ngVLA beam is given in the bottom left of the middle panel. The simulated ngVLA observations have a 1 hr on-source $\sigma_{\rm RMS} = 0.48~\mu \rm Jy~beam^{-1}$ and 1000 km baselines.}
\end{figure}

The ngVLA will be able to spatially resolve a very high level of detail in the inner regions of disks. To illustrate this, Fig.\,1 shows simulated observations of a detailed disk model with both the ngVLA and ALMA. The model shown is a global hydrodynamic simulation of a protoplanetary disk in the thin-disk limit, starting with the collapse of molecular cloud \citep{VorobyovBasu2015}. The model used canonical values of layered disk parameters (e.g., ${\rm T_{crit}=1100 K, \Sigma_{crit}=100 g/cm^2}$) and reproduced eruptions very similar to the observations of FUor outbursts. 

To simulate this disk model with the ngVLA, we assume the disk emission is consistent with 100 GHz thermal continuum (i.e., it does not include any non-thermal effects, disk winds, or ionized emission), the system is at a distance of 150 pc, and the disk flux is normalized to 35 mJy. The highest resolution antenna configuration of 0.7 mas was used and the background noise is set to $0.48~\mu \rm Jy~ beam^{-1}$ \citep[achievable in 1 hr on-source in ngVLA Band 6;][]{selina18}. For comparison to ALMA, a simulated observation of the same disk is included with a sensitivity equivalent to 4 hours on-source (the most extended ALMA baseline of $\sim16$ km corresponds to a resolution of $\sim0.04$ arcsec at 100 GHz).

The $10\times10$ au region of the disk model shows the complexity of the inner accretion disk which can be resolved by ngVLA, but not with current capabilities of ALMA. Should significant optically thick disk winds or non-thermal emission be present in a system at 100 GHz, then it could indeed make resolving the disk structure more complicated. The model in Fig.\,1 predicts that this inner region is very dynamic and will show detectable changes in features on the timescale of a few years. The perturbations in the disk during the outburst propagate outwards, producing spirals or concentric rings also observed in young stellar disks. 

ALMA is the current state-of-the-art facility in regards to high sensitivity and resolution observations of EX/FUor-types stars. As ALMA is located in the Southern hemisphere, there are potentially many targets in the Northern hemisphere that are not easily detectable or resolved with the current Northern hemisphere observing facilities (e.g., SMA and NOEMA). Therefore, the ngVLA is uniquely positioned to provide detailed observations of a large fraction of the sky. 

\section{Synergies at Other Wavelengths}

The frequency in which EX/FUor-types of outbursts occur is not well constrained due to the relatively small number of outburst events. Determining the frequency of these events requires large scale surveys, which is more practical with optical, near-IR, and IR telescopes. Future optical/IR survey facilities, such as LSST and WFIRST, will likely identify new outbursts and the ngVLA will be a key tool in follow-up observations and an accurate characterization of the systems. The SKA may be able to detect non-thermal emission in EX/FUors and will provide complimentary longer wavelength data.

Higher-energy observations, e.g., in the X-ray regime, can provide important complementary information on plasma heating in magnetic fields (through corona-type emission like in active stars) and the presence of {\it neutral} gas components in the line of sight, potentially indicating the presence of neutral winds or accretion flows \citep{liebhart14}. Clarifying the role of X-rays generated during massive outbursts in the evolution and dispersal of planet-forming disks and envelopes is essential to progress in connecting the first stages of planet formation to mature planetary systems.

While many of the morphological details cannot be revealed with current facilities, mm telescopes such as ALMA and NOEMA can still determine many of the bulk properties of a given EX/FUor. For example, ALMA can provide the total $\sim$mm flux which will be useful for spectral index constraints in optically thin regimes. There are also many aspects of disk chemistry that can be measured with ALMA (e.g., CO(3-2) at 345 GHz) and future IR facilities such as JWST, that will be complimentary to the disk chemistry information obtained with the ngVLA.

The ngVLA will provide unprecedented spatial resolution and sensitivity in the $1-116$ GHz frequency and contribute to the broad spectral coverage that is imperative to accurately models of EX/FUors and young outbursting systems.




\end{document}